\documentclass[aps,prd,twocolumn,superscriptaddress,nofootinbib]{revtex4-1}

\usepackage{latexsym}
\usepackage{amsmath}
\usepackage{amssymb}
\usepackage{amsfonts}
\usepackage{slashed}

\usepackage{color}

\usepackage{supertabular} 
\usepackage{placeins}
\usepackage{epsfig}
\usepackage{graphicx}

\usepackage{color,soul}

\usepackage[breaklinks,pageanchor]{hyperref}

\hypersetup{
  colorlinks=true,
  linkcolor=blue,
  filecolor=magenta,
  urlcolor=blue,
  citecolor=blue,
}

\begin{document}


\title{Study of the $\Omega_{ccc}\Omega_{ccc}$ and $\Omega_{bbb}\Omega_{bbb}$ dibaryons  in constituent quark model}

\author{Pablo Mart\'\i n-Higueras}
\email[]{pablo.higueras@alu.uhu.es}
\affiliation{Departamento de Ciencias Integradas y Centro de Estudios Avanzados en F\'\i sica,
Matem\'atica y Computaci\'on, Universidad de Huelva, 21071 Huelva, Spain}

\author{David R. Entem}
\email[]{entem@usal.es}
\affiliation{Grupo de F\'isica Nuclear, Universidad de Salamanca, E-37008 Salamanca, Spain}
\affiliation{Instituto Universitario de F\'isica Fundamental y Matem\'aticas (IUFFyM), Universidad de Salamanca, E-37008 Salamanca, Spain}

\author{Pablo G. Ortega}
\email[]{pgortega@usal.es}
\affiliation{Instituto Universitario de F\'isica Fundamental y Matem\'aticas (IUFFyM), Universidad de Salamanca, E-37008 Salamanca, Spain}

\author{Jorge Segovia}
\email[]{jsegovia@upo.es}
\affiliation{Departamento de Sistemas F\'isicos, Qu\'imicos y Naturales, \\ Universidad Pablo de Olavide, E-41013 Sevilla, Spain}

\author{Francisco Fern\'andez}
\email[]{fdz@usal.es}
\affiliation{Grupo de F\'isica Nuclear, Universidad de Salamanca, E-37008 Salamanca, Spain}
\affiliation{Instituto Universitario de F\'isica Fundamental y Matem\'aticas (IUFFyM), Universidad de Salamanca, E-37008 Salamanca, Spain}

\date{\today}

\begin{abstract}
Dibaryons are the simplest system in which the baryon-baryon interaction, and hence the underlying quark-quark interaction, can be studied in a clear way.
Although the only dibaryon known today is the deuteron (and possibly the $d^*$), fully heavy dibaryons are good candidates for bound states because in such systems the kinetic energy is small and the high symmetry of the wave function favours binding.
In this study, the possible existence of $\Omega_{ccc}\Omega_{ccc}$ and $\Omega_{bbb}\Omega_{bbb}$ dibaryons is investigated in the framework of a constituent quark model that satisfactorily describes the deuteron, the $d^*(2380)$ and the $NN$ interaction. $J^P=0^+$ candidates are found in both systems with binding energies of the order of MeV.
\end{abstract}



\maketitle


\section{Introduction}
\label{sec:intro}

Understanding the nucleon-nucleon interaction has been one of the priority problems in Nuclear Physics since Yukawa's one pion exchange theory. The subsequent development of QCD paved the way to describe the strong interactions in terms of quark degrees of freedom and facilitate to enlarge the field to other flavors like charm en bottom.

Dibaryons are the simplest systems in which these studies can be addressed in a transparent way.
Until recently, the only well-established bound state of two baryons was the deuteron. Then, in 2011, another unstable light dibaryon, the $d^*(2380)$, was reported by the WASA-at-COSY Collaboration~\cite{WASA-at-COSY:2011bjg} from the double pionic fusion reaction $pn\to d\pi^0\pi^0$. This resonance can be described as a nonstrange $\Delta\Delta$ dibaryon with $I(J^P)=0(3^+)$. In 1989, Goldman noted that due to the special symmetry of such a state, any model based on confinement and gluon exchange should predict it~\cite{Goldman:1989zj}. The long history of the search for dibaryons in the light quark sector can be found in Ref~\cite{Clement:2016vnl}.

It is well known that the binding of the deuteron is due to the coupling of the $^3S_1$ and $^3D_1$ partial waves by one-pion exchange tensor interactions. Similarly, the binding of the $d^*(2380)$ can be explained in terms of Goldstone-boson exchanges~\cite{PhysRevC.89.034001}. These two systems then prove that the interaction binding these dibaryons arises from QCD chiral symmetry breaking in the light quark sector.

Another interesting system is the fully heavy dibaryon. In such a system the relativistic effects are negligible and the kinetic energy is small. As originally pointed out by Bjorken~\cite{Bjorken:1985ei}, the triply-charmed baryon $\Omega_{ccc}$ is stable against strong interactions. This fact opens the possibility to study systems like $\Omega_{ccc}\Omega_{ccc}$ or $\Omega_{bbb}\Omega_{bbb}$. Moreover, in contrast to the deuteron and the $d^*$ case, the latter systems provide an ideal scenario to explore the baryon-baryon interaction in an environment free of chiral dynamics.

In this work we will focus on the study of the fully heavy dibaryons. Two recent Lattice QCD calculations have explored these systems: Ref.~\cite{Lyu_2021} showed that $\Omega_{ccc}\Omega_{ccc}$ is loosely bound by $5.68(0.77)$ MeV, while Ref.~\cite{Mathur_2023} found a very deep $\Omega_{bbb}\Omega_{bbb}$ state with a binding energy of $81_{-16}^{+14}$ MeV. These conclusions are confirmed by several quark model calculations, but are contradicted by others. For example, Huang {\it et al.}~\cite{Huang:2020bmb}, using a constituent quark model based on the one-gluon exchange interaction and the resonating group method, studied the possible bound states of the $\Omega_{ccc}\Omega_{ccc}$ and $\Omega_{bbb}\Omega_{bbb}$, among others. They found a $J^P=0^+$ bound state for the $\Omega_{ccc}\Omega_{ccc}$ system with a binding energy of $2.5$ MeV and another $\Omega_{bbb}\Omega_{bbb}$ state bound by $0.9$ MeV, contrary to naive espectations. Deng~\cite{Deng_2023} performed a study of the di-$\Delta^{++}$, di-$\Omega_{ccc}$ and di-$\Omega_{bbb}$ systems using a naive one-gluon exchange quark model and a chiral quark model including $\pi$ and $\sigma$ exchanges between quarks.
Obviously, in this case these parts of the interaction apply only to the light quarks, but the set of parameters is different in the two models. Both studies predict very shallow di-$\Omega_{ccc}$ and di-$\Omega_{bbb}$ states with binding energies around $1$ MeV.
Using a different model, namely QCD sum rules, Wang~\cite{Wang_2022} found for each di-$\Omega_{ccc}$ and di-$\Omega_{bbb}$ systems two
$J^P=0^+$ and $J^P=1^-$ states that are slightly below their respective thresholds.

On the other hand, several studies within the quark model have ruled out the existence of fully heavy dibaryons. In Ref~\cite{Richard:2020zxb} the authors investigated the existence of $bbbccc$ dibaryons and extrapolated their results to the properties of the $bbbbbb$ and $cccccc$ systems. They found no bound states for $\Omega_{ccc}\Omega_{ccc}$ or $\Omega_{bbb}\Omega_{bbb}$ combinations. On the other hand, Alcaraz-Peregrina {\it et al.}~\cite{Alcaraz_Pelegrina_2022} used the Difussion Montecarlo technique to describe fully heavy compact six-quark arrangements. They found that all the hexaquarks have smaller masses than those of their constituents, i.e., all the hexaquarks are bound systems.
However, their masses are also larger than those of any pair of baryons into which they can be divided. This means that each hexaquark is unstable with respect to its splitting into two baryons. Finally, two more calculations, in the framework of the constituent quark model~\cite{Lu:2022myk} or the extended chromomagnetic model~\cite{Weng_2024}, showed that all the fully heavy dibaryons lie above their corresponding baryon-baryon thresholds.

In view of this controversial situation, since different approaches lead to quite different conclusions, we will study the possible existence of $\Omega_{ccc}\Omega_{ccc}$ and $\Omega_{bbb}\Omega_{bbb}$ dibaryons using the constituent quark model of Ref.~\cite{Fernandez:1993hx} and its extension to the heavy quark sector~\cite{Vijande:2004he,PhysRevD.78.114033}, which has been able to describe a large variety of hadronic phenomenology. In particular, the model reproduces the properties of the deuteron~\cite{PhysRevC.50.2246,PhysRevC.66.047002} and predicts the existence of the $d^*(2380)$ as a $\Delta\Delta$ dibaryon~\cite{PhysRevC.56.84,Valcarce:2001in}. Although the binding energy of the $d^*$ predicted in the latter references is smaller than the experimental value, it is laso worth mentioning that the these studies were performed without coupling to the $NN$ channel. 

The paper is structured as follows. In Sec.~\ref{sec:theory} we describe the main aspects of our theoretical model, giving details about the wave functions used to describe $\Omega_{ccc}$ ($\Omega_{bbb}$) baryons and the way we derive the $\Omega_{ccc}\Omega_{ccc}$ interaction using the Resonating Group Method (RGM). Section~\ref{sec:results} is devoted to presenting our results for the possible dibaryons. Finally, we summarize and give some conclusions in Sec.~\ref{sec:epilogue}.

\section{Theoretical formalism}
\label{sec:theory} 
\subsection{The constituent quark model}
\label{subsec:quarkmodel}

Our theoretical framework is a QCD-inspired constituent quark model (CQM) proposed in Ref.~\cite{Vijande:2004he} and extended to the heavy quark sector in Ref.~\cite{PhysRevD.78.114033}.
The main pieces of the model are the constituent light quark masses  and Goldstone-boson exchanges, which appears as consequences of spontaneous chiral symmetry breaking of the QCD Lagrangian together with perturbative one-gluon exchange (OGE) and nonperturbative color confining interactions.

Following Diakonov~\cite{Diakonov:2002fq}, a simple Lagrangian invariant under chiral transformations can be written as
\begin{equation}
{\mathcal L} = \bar{\psi}(i\, {\slash\!\!\! \partial} - M(q^{2}) U^{\gamma_{5}}) \,\psi \,,
\end{equation}
where $M(q^2)$ is the dynamical (constituent) quark mass and $U^{\gamma_5} = e^{i\lambda _{a}\phi ^{a}\gamma _{5}/f_{\pi}}$ is the matrix of Goldstone-boson fields that can be expanded as
\begin{equation}
U^{\gamma _{5}} = 1 + \frac{i}{f_{\pi}} \gamma^{5} \lambda^{a} \pi^{a} - \frac{1}{2f_{\pi}^{2}} \pi^{a} \pi^{a} + \ldots
\end{equation}
The first term of the expansion generates the constituent quark mass, while the second term gives rise to a one-boson exchange interaction between quarks. The main contribution of the third term comes from the two-pion exchange which has been simulated by means of a scalar-meson exchange potential.

In the heavy quark sector, chiral symmetry is explicitly broken and Goldstone-boson exchange does not occur. However, the full interaction constrains the model parameters through the light-meson phenomenology~\cite{Vijande:2004he,PhysRevD.78.114033}. Thus, OGE and confinement are the only remaining interactions between the heavy quarks.

The OGE potential is generated from the vertex Lagrangian
\begin{equation}
{\mathcal L}_{qqg} = i\sqrt{4\pi\alpha_{s}} \, \bar{\psi} \gamma_{\mu} G^{\mu}_{c} \lambda^{c} \psi,
\label{Lqqg}
\end{equation}
where $\lambda^{c}$ are the $SU(3)$ colour matrices, $G^{\mu}_{c}$ is the gluon field and $\alpha_{s}$ is the strong coupling constant. The scale dependence of $\alpha_{s}$ allows a consistent description of light, strange and heavy mesons. Its explicit expression can be found in, {\it e.g.}, Ref.~\cite{Vijande:2004he},

\begin{equation}
  \alpha_s(\mu)=\frac{\alpha_0}{\ln \left(\frac{\mu^2+\mu_0^2}{\Lambda_0^2}\right)}
\end{equation}

Regarding the confinement potential, it is well known that multi-gluon exchanges produce an attractive linearly rising potential proportional to the distance between infinite-heavy quarks~\cite{Mateu:2018zym}. However, sea quarks are also important components of the strong interaction dynamics that contribute to the screening of the rising potential at low momenta and eventually to the breaking of the quark-antiquark binding string~\cite{Bali:2005fu}. Our model tries to mimic this behaviour with a screening potential at high distances.

Then, the full interaction between heavy quarks is given by

\begin{align}
  V_{ij}(r) &= \bigg[ -a_c(1-e^{-\mu_c r}) + \Delta + \frac{\alpha_s(\mu)}{4} \frac{1} {r} \bigg] (\vec \lambda_i \cdot \vec \lambda_j) \nonumber
\\
V_{ij}^S(r) &= -\frac{\alpha_s(\mu)}{4} \frac{1}{6m_im_j}\frac{e^{-r/r_0(\mu)}}{rr_0^2(\mu)} (\vec \sigma_i \cdot \vec \sigma_j)
(\vec \lambda_i \cdot \vec \lambda_j)\nonumber
\\
V_{ij}^T(r) &= -\frac{1}{16} \frac{\alpha_s(\mu)}{m_im_j} S_{ij}\,(\vec \lambda_i \cdot \vec \lambda_j)\times\nonumber\\
&\times \bigg[\frac{1}{r^3}-\frac{e^{-r/r_g(\mu)}}{r}\bigg (\frac{1}{r^2}+
\frac{1}{3r_g^2(\mu)}+\frac{1}{rr_g(\mu)}\bigg )\bigg ]
\end{align}
where $r_0(\mu)=\hat r_0 \frac{m_n}{2\mu}$ with $\mu$ the reduced mass of the $(ij)$ heavy quark pair, $\vec\lambda$ are the colour matrices, $\vec\sigma$ the spin matrices and $S_{ij}=3(\vec \sigma_i\cdot\hat r)(\vec \sigma_j\cdot\hat r)-(\vec\sigma_i\cdot \vec\sigma_j)$ the tensor operator of the $(ij)$ pair with $\vec r$ their relative position.

All the parameters of the model are given in Table~\ref{param}. We have not included the spin-orbit interaction parts coming from the one-gluon exchange and confinement because they should give small contributions in this calculation. For the same reason, the spin-tensor terms are neglected in the calculation of the $\Omega_{ccc}$ ($\Omega_{bbb}$) masses, but are included in the $\Omega_{ccc}$$\Omega_{ccc}$ ($\Omega_{bbb}$$\Omega_{bbb}$) interaction.

\begin{table}
\begin{center}
  \begin{tabular}{ccc}
    \hline
    \hline
    Quark masses (MeV) & $m_c$ & 1763 \\
                       & $m_b$ & 5110 \\
    \hline
    OGE                & $\hat r_0$ (fm)                          & 0.181    \\
                       & $\alpha_0$                               & 2.118    \\
                       & $\Lambda_0$ (fm$^{-1}$)                  & 0.113    \\
                       & $\mu_0$ (MeV)                            & 36.976   \\
    \hline
    Confinement        & $a_c$ (MeV)                               & 507.4    \\
                       & $\mu_c$ (fm$^{-1}$)                      & 0.576    \\
                       & $\Delta$ (MeV)                           & 184.432  \\
    \hline
    \hline
\end{tabular}
\caption{\label{param} Parameters for the quark-quark interaction.}
\end{center}
\end{table}

\subsection{The wave function of the $\Omega_{ccc}$($\Omega_{bbb}$)}
\label{subsec:wave functions}

A precise definition of the wave functions of the $\Omega_{ccc}$ and $\Omega_{bbb}$ baryons (henceforth $\Omega_{QQQ}$) is an essential part of the calculation,  because it defines the size of the baryon, which is important for the baryon-baryon interaction. 

Once we know the quark-quark interaction, the $\Omega_{QQQ}$ wave function can be calculated by solving the Schr\"odinger equation with the Gaussian Expansion Method (GEM)~\cite{Hiyama:2003cu}. In the GEM framework one makes an expansion in gaussian wave functions but instead of using only one
set of Jacobi coordinates, one includes the lowest orbital angular momentum wave functions using the three sets of possible Jacobi coordinates.
The reason to use different sets is that lowest angular momentum wave functions in one set generates higher angular momentum wave functions
in the other sets, making a very numerically efficient way to include such high angular momentum components.

However the wave function given by GEM would be quite complicate and would make the calculation of the dibaryon interaction slow. Alternatively, for the calculation of the dibaryon interaction (that would be justified later), the following orbital wave function can be used
\begin{align}\label{eq:baryonWF}
	\phi(\vec p_{\xi_1},\vec p_{\xi_2}) &=
	\bigg[ \frac{2b^2}{\pi} \bigg]^{3/4}  e^{-b^2 p_{\xi_1}^2}
	\bigg[ \frac{3b^2}{2\pi} \bigg]^{3/4}  e^{-\frac{3b^2}{4} p_{\xi_1}^2}
\end{align}
where $p_{\xi_i}$ are the Jacobi coordinates defined as
\begin{align}
	\vec p_{\xi_1} &= \frac{1}{2} (\vec p_1 - \vec p_2)
	\nonumber \\ 
	\vec p_{\xi_2} &= \frac 2 3 \vec p_3 - \frac 1 3 (\vec p_1 + \vec p_2)
\end{align}
In the notation we use for the baryon calculation this corresponds to mode 3 of the GEM basis 
using only one gaussian with angular momentum zero and the
parameters 
$\nu=\frac{1}{4b^2}$ and $\lambda=\frac{1}{3b^2}$. Notice that fixing the relation between the parameters of the gaussians $\nu$ and $\lambda$
to these values ($\nu=\frac 3 4 \lambda$), the orbital wave functions is totally symetric, necessary to get a totally antisymetric
wave function for the baryon of lowest energy.
The spin wave function has to be also symmetric and implies $S=\frac 3 2$ and the color wave function will be a color singlet.

So our wave function for the baryon is
\begin{align}
	\psi_B &= \phi(\vec p_{\xi_1},\vec p_{\xi_2}) \chi_B \xi_c[1^3]
\end{align}
with $\chi_B=((\frac{1}{2} \frac{1}{2}) 1 \frac{1}{2}) \frac{3}{2})$ the spin wave function and $\xi_c[1^3]$ a singlet color wave function.

Using the wave function of Eq.~\eqref{eq:baryonWF}, the kinetic energy is given by
\begin{align}
	T &= \langle \psi_B | \frac{p_{\xi_1}^2}{m} + \frac{3p_{\xi_2}^2}{4m} | \psi_B \rangle = \frac{3}{2m b^2}
\end{align}
For the interaction energy we can evaluate $\langle \psi_B | V_{12} | \psi_B \rangle$ and multiply by 3, since we have
3 interactions between equivalent quarks. It is easier to evaluate it in coordinate space. The wave function in coordinate space is
\begin{align}
	\phi_B(r_3,R_3) = 
	\bigg[ \frac{1}{2\pi b^2} \bigg]^{3/4} e^{-\frac{r_3^2}{4b^2}}
	\bigg[ \frac{2}{3\pi b^2} \bigg]^{3/4} e^{-\frac{R_3^2}{3b^2}}
\end{align}
and so
\begin{align}
	\langle \psi_B | V_{12} | \psi_B \rangle &= 4\pi \bigg[ \frac{1}{2\pi b^2} \bigg]^{3/2}
	\int_0^\infty r_3^2 dr_3 e^{-\frac{r_3^2}{2b^2}} V(r_3)
\end{align}

The mean value distance between quarks is given by
\begin{align}
	\sqrt{\langle r_{ij}^2 \rangle} &= \sqrt 3 b
\end{align}

and the mass is given by
\begin{align}
	M = 3m_b + T + 3 \langle \psi_B | V_{12} | \psi_B \rangle
\end{align}

Finally, the value of the $b$ parameter is obtained by minimizing the mass

\begin {align}
\frac{\partial M}{\partial b}=0
\end{align}

\begin{table}
 \begin{tabular}{r|rr|rr}
 \hline\hline
  & \multicolumn{2}{c}{$\Omega_{ccc}$} & \multicolumn{2}{c}{$\Omega_{bbb}$} \\ \hline
  & $\partial M/\partial b$ & GEM & $\partial M/\partial b$ & GEM \\\hline
$M$ [MeV] & 4810.9 & 4798.6 & 14413.8 & 14396.9\\
$\sqrt{\langle r_{ij}^2 \rangle}$ [fm]& 0.4360 & 0.4432 & 0.2716 & 0.2762\\
$\langle T \rangle$ [MeV]& 522.9 & 522.8 & 465.0 & 471.5\\
$\langle V  \rangle$ [MeV]& -333.7 & -337.7& -460.4 & -468.2\\
\hline\hline
 \end{tabular}
 \caption{\label{tab2} Parameters for the $\Omega_{ccc}$ and $\Omega_{bbb}$ baryons obtained from the mass minimization procedure ($\frac{\partial M}{\partial b}=0$) and the Gaussian Expansion Method.}
\end{table}

Although the GEM method provides a more complete description of the wave function as mentioned before, 
the calculation is simplified if we use the analytical wave function of Eq.~\eqref{eq:baryonWF}. 
In Table~\ref{tab2} we show the results of the $\Omega_{ccc}$ and $\Omega_{bbb}$ wave functions using the mass minimization procedure and 
compare with the GEM solution to justify the use of the simple wave function given by Eq.~\eqref{eq:baryonWF}.
We see that we get a resonable agreement for the sizes and energies in both cases, although the
agreement is better in the beauty sector. The minimal values for $b$ are given by
$b_{\rm min}=0.15679$ fm for the $\Omega_{bbb}$ and
$b_{\rm min}=0.25172$ fm for the $\Omega_{ccc}$.

\section{The $\Omega_{QQQ}\Omega_{QQQ}$ interaction}

The system under study has six identical quarks. Then, the baryon-baryon total wave function must be fully antisymmetric.
As the wave function of the baryons is already antisymmetric, the antisymmetrizer operator is just given by,
\begin{align}
	{\mathcal A} = 1-9P_{36}
\end{align}

In order to obtain the effective baryon-baryon interaction from the underlying quark dynamics we use the Resonating Group Method (RGM)~\cite{Wheeler:1937zza, Tang:1978zz}. Then, we need to solve the projected Schr\"odinger equation,
\begin{align}
	0=\bigg( \frac{p'^2}{2\mu_{\Omega\Omega}} -E \bigg) \chi(\vec P')
	+\int& \bigg( \,^{\rm RGM}V_D(\vec P',\vec P_i) +\nonumber\\
	&+\,^{\rm RGM}K(\vec P',\vec P_i) \bigg)\chi(\vec P_i) d^3 P_i
\end{align}
where $\vec P'$ ($\vec P_i$) is the relative $\Omega_{QQQ}-\Omega_{QQQ}$ final (initial) momentum, 
$E=E_T-2M_{\Omega}$ the relative energy of the system with respect to the threshold, 
$^{\rm RGM}V_D(\vec P',\vec P_i)$
is the direct kernel and $^{\rm RGM}K(\vec P',\vec P_i)$ is the exchange kernel and
$\mu_{\Omega\Omega}$ is 
the reduced mass of two $\Omega_{QQQ}$ baryons. 

Here, $M_{\Omega}$ is
\begin{align}
	M_{\Omega} &= 3 m_b + \frac{3}{2m_Q b^2} + 3 E_{int}
	\\
	E_{int} &= \langle V_{ij} \rangle = \int d^3 q e^{-\frac{q^2b^2}{2}} \langle V_{ij}(q) \rangle
\end{align}

The direct term will be zero in the present model since the color coefficients, $(\vec \lambda_i\cdot \vec \lambda_j)$, are zero between color singlets.

Then, the full interaction is driven by exchange diagrams, which take into account the quark rearrangement between baryons. The exchange kernel can be written as,

\begin{align}
\,^{\rm RGM}K(\vec P',\vec P_i) =& \,^{\rm RGM}T(\vec P',\vec P_i) +
^{\rm RGM}V_{ij\,E} (\vec{P}' , \vec{P}_i ) - \nonumber\\
&-E_T \,^{\rm RGM}N(\vec P',\vec P_i)
\end{align}
where $^{\rm RGM}T(\vec P',\vec P_i)$ is the exchange kinetic term, $^{\rm RGM}N(\vec P',\vec P_i)$ is 
a normalization term and $^{\rm RGM}V_{ij\,E}(\vec P',\vec P_i)$ is the exchange potential 
(for explicit expressions see, e.g., Refs.~\cite{Ortega:2022tbl,Fernandez:2019ses}).

\section{Results}
\label{sec:results} 

Let us first study the $\Omega_{bbb}\Omega_{bbb}$ system. 
One of the states in $S$ wave is the $J^P=0^+$, which corresponds to the 
$^1S_0$ and $^5 D_0$ partial waves. As in the case of the deuteron, $S$ and $D$ waves are mixed. 
We first calculate the binding energy considering the parameter $b$ and the reduced mass given
by the minimization procedure. 
Without tensor interactions they are decoupled and only a bound state appears in the $^1S_0$ partial 
wave. The binding energy of this state is
$E=-1.9859$ MeV. The $^5 D_0$ partial wave is not bound. If we include the tensor interaction of OGE 
the partial waves are coupled and the binding energy
increases very slightly to $E=-1.9876$ MeV. The probability of the $D$ wave is only $6.6\cdot 10^{-4}\%$. As in the deuteron, the binding
energy has a sizeable cancellation between the kinetic and interaction parts. The mean value of the kinetic energy
is $\langle T \rangle = 11.2$ MeV, while for the interaction we have $\langle V \rangle = -13.2$ MeV. 
The confinement interaction dominates and gives the needed attraction to bind the system. If we exclude the OGE we get $E=-7.3698$ MeV with
$\langle T \rangle = 20.9$ MeV and $\langle V \rangle = -28.3$ MeV.

\begin{figure}[t]
\begin{center}
	\scalebox{0.80}{\includegraphics{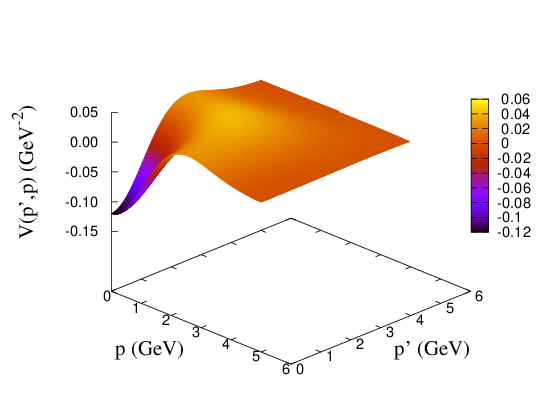}}
\caption{\label{Pot} The potential $V(p',p)$ in the $^1S_0$ partial wave for the $\Omega_{bbb}\Omega_{bbb}$ interaction.}
\end{center}
\end{figure}

\begin{figure}[t]
\begin{center}
	\scalebox{0.7}{\includegraphics{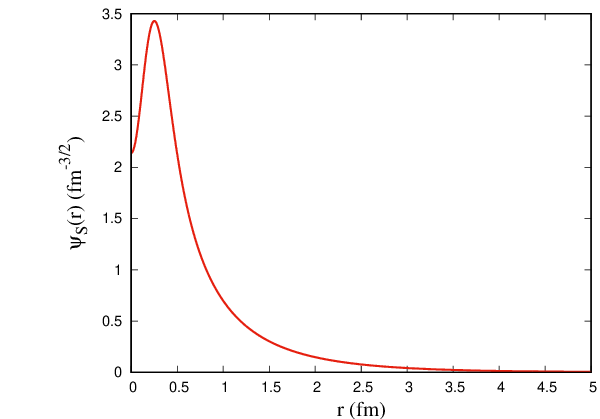}}
	\scalebox{0.7}{\includegraphics{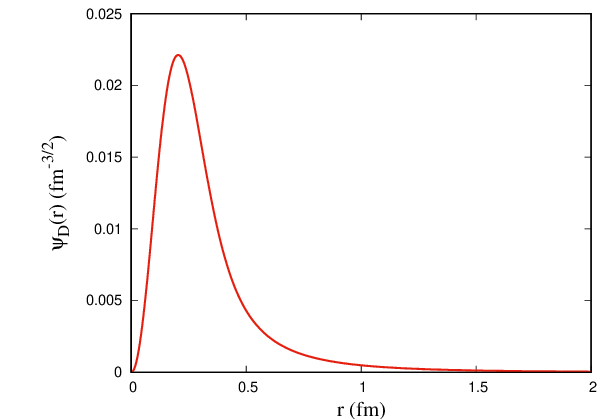}}
\caption{\label{Psi} The relative wave functions in the $^1S_0$ and $^5D_0$ partial waves for the $\Omega_{bbb}\Omega_{bbb}$ dibaryon.}
\end{center}
\end{figure}

The potential for the $^1S_0$ partial wave is given in Fig.~\ref{Pot}. The relative wave functions are shown in Fig.~\ref{Psi}.

If we consider $b=\frac{\sqrt{r_{ij}^2}}{\sqrt 3}$ and the reduced mass given by the GEM calculation
we get a binding energy in the coupled case of $E=-1.81$ MeV. With $b$ given by the minimization 
procedure and the reduced mass is given by GEM we get $=-1.9754$ MeV. The effect of 
the different reduced mass
is very small and dominates the effect of the different $b$ parameters. In principle with only one
gaussian one should use the value given by the minimization procedure, but this gives us a feeling
of the uncertainty due to the simplification of the wave function. Although the binding energy
varies a little bit, in both cases the system is bounded.

Another possible state is the $J^P=2^+$ which includes the $^5S_2$, $^1D_2$, $^5D_2$ and $^5G_2$ partial waves. None of them
are bound. One could expect a bound state for the $^5S_2$ partial wave, but in this case one can see that the potential coming
from the $\vec \lambda_i \cdot \vec \lambda_j$ have opposite sign for $S=2$ with respect to $S=0$. So if we have attraction for the $S=0$,
this implies repulsion for $S=2$. Higher partial waves are more difficult to bind.

Antisymmetry implies $L+S=$~even and parity is given by $P=(-1)^L$. So for $P=+$, the spin $S$ has to be even. This means that $1^+$ and $3^+$ can
be only in $D$ or $G$ waves, which will be difficult to bind as it was seen for the $0^+$ and $2^+$ states. In more detail:
\begin{itemize}
\item We start with the $1^+$ state and include $^5 D_1$ which is the only partial wave. It should be the same as the $^5D_2$ partial wave
with the exception of the contribution of the OGE tensor interaction. It does not bind.
\item For the $3^+$ state we have the $^5 D_3$ and $^5G_3$ partial waves and they do not bind.
\end{itemize}

\begin{figure}[t]
\begin{center}
	\scalebox{0.7}{\includegraphics{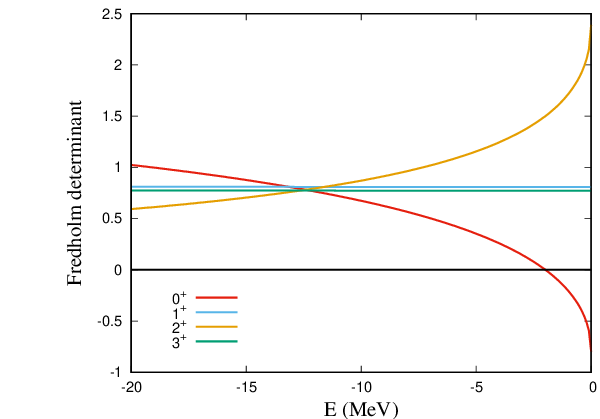}}
\caption{\label{Fredholm} The Fredholm determinant of the $\Omega_{bbb}\Omega_{bbb}$ system for the $0^+$, $1^+$, $2^+$ and $3^+$ channels. Only the $0^+$ crosses the zero.}
\end{center}
\end{figure}

We give in Fig.~\ref{Fredholm} the Fredholm determinant for the 4 different $J^+$ quantum numbers, where we can see that only
the $0^+$ channel binds.

\begin{figure}[t]
\begin{center}
	\scalebox{0.7}{\includegraphics{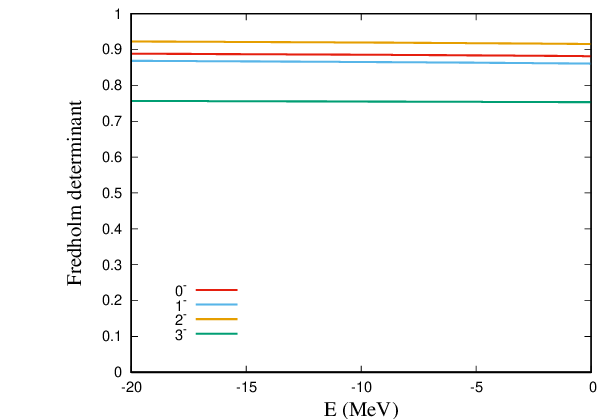}}
\caption{\label{Fredholmb} The Fredholm determinant of the $\Omega_{bbb}\Omega_{bbb}$ system  for the $0^-$, $1^-$, $2^-$ and $3^-$ channels. None of them crosses the zero.}
\end{center}
\end{figure}

Regarding possible $P=-$ states, this would imply odd partial waves and odd total spin. We have analyzed the $J^P=\{0^-,1^-,2^-,3^-\}$, finding no additional bound states.
Again, in Fig.~\ref{Fredholmb} the Fredholm determinant for the 4 different $J^-$ quantum numbers is shown, where we can see that no bound state is predicted.

Concerning the $\Omega_{ccc}\Omega_{ccc}$ system, the situation is similar to the 
$\Omega_{bbb}\Omega_{bbb}$ system, and we only find a bound state in the $0^+$ channel. 
The binding energy is $E=-0.7104$ MeV with a $D$-state probability of $1.7\cdot 10^{-3}\%$. The mean values of kinetic and interaction terms are $\langle T \rangle = 7.46$ MeV and $\langle V \rangle = -8.17$ MeV.
In this case we used the $b$ parameter from the minimization procedure and the reduced mass from the GEM.
Using the reduced mass from the minimization parameter the binding energy changes to $E=-0.7288$ MeV
and both parameters from the GEM to $E=-0.62$ MeV.

\subsection{Dependence on the model parameters}

We analyze the dependence on the parameters of the model for the $J^P=0^+$ state to see in which parameter space region
the system will not bind. In all cases we use the minimization procedure to obtain $b$ and
$\mu_{\Omega\Omega}$.

\begin{figure}[t]
\begin{center}
        \scalebox{0.7}{\includegraphics{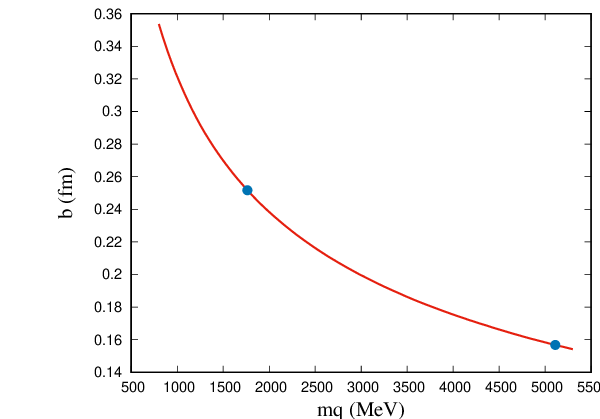}}
        \scalebox{0.7}{\includegraphics{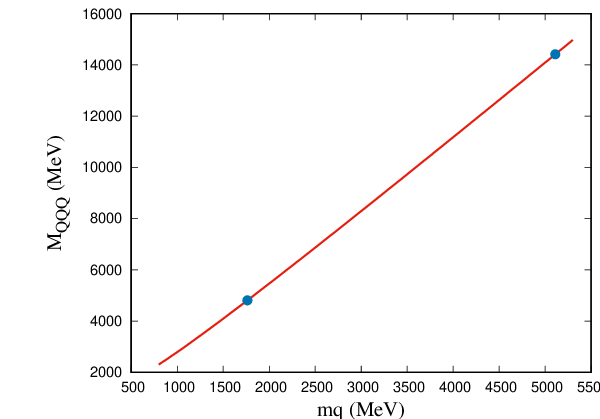}}
        \scalebox{0.7}{\includegraphics{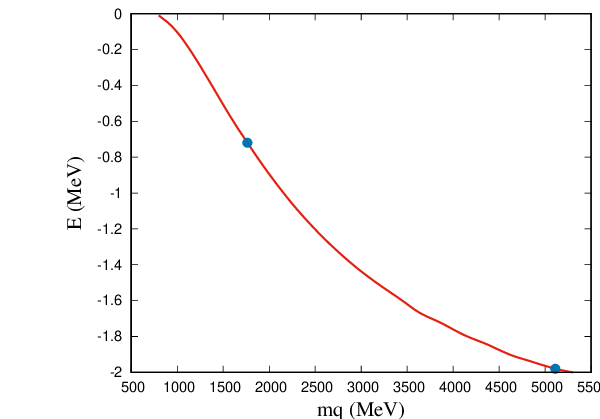}}
        \caption{\label{mq} Quark mass dependence of $b$, $M_{QQQ}$ and $E$. We show by a dot the result of the Chiral Quark Model.}
\end{center}
\end{figure}

The dependence on the quark mass $m_q$ is shown in Fig.~\ref{mq}. Notice that some of the parameters
of the potential depends on $m_q$ since we use scale dependent parameters. We see that the system binds
reducing the quark mass up to $m_q\sim 800-900$ MeV.

Our model has an effective string tension given by
\begin{eqnarray} 
        \sigma = \frac{8}{3} a_c \mu_c = 0.1537\,{\rm GeV}^2
\end{eqnarray}

\begin{figure*}[t]
\begin{center}
        \scalebox{0.7}{\includegraphics{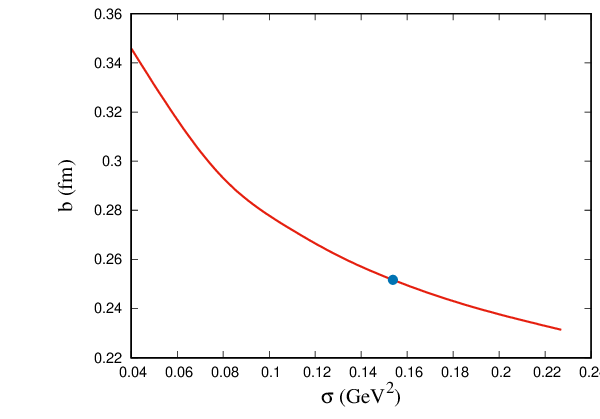}}
        \scalebox{0.7}{\includegraphics{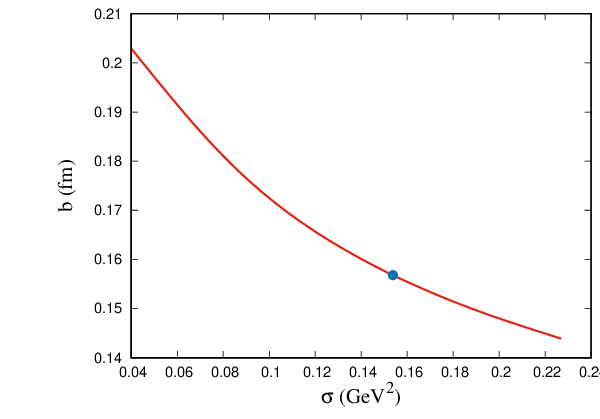}}
        \scalebox{0.7}{\includegraphics{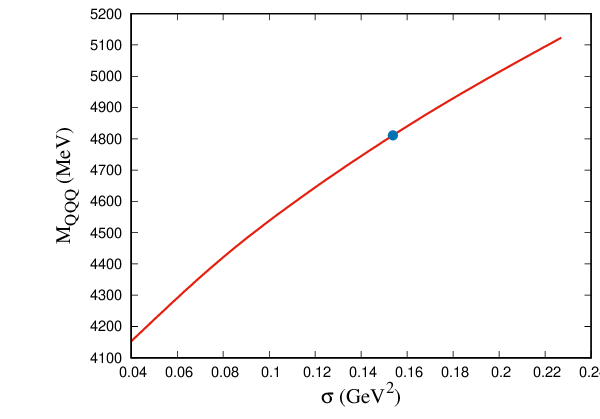}}
        \scalebox{0.7}{\includegraphics{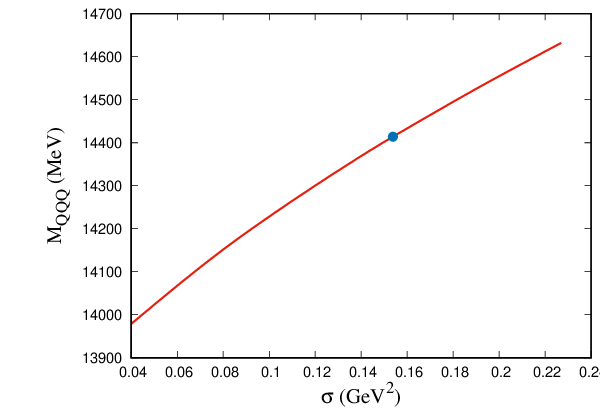}}
        \scalebox{0.7}{\includegraphics{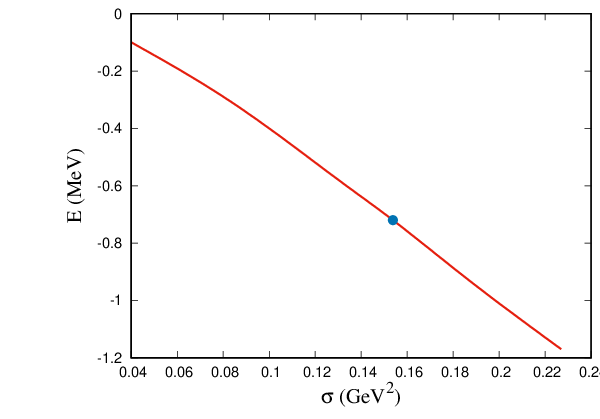}}
        \scalebox{0.7}{\includegraphics{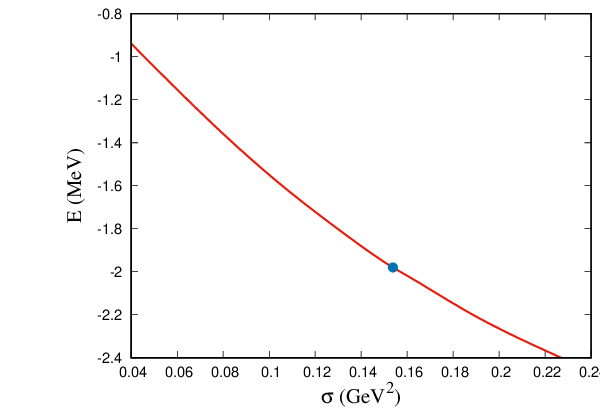}}
        \caption{\label{sigma} Dependence of $b$, $M_{QQQ}$ and $E$ on the effective string tension $\sigma= \frac{8}{3} a_c \mu_c$ for
$m_q=m_c$ (left) and $m_q=m_b$ (right). We show by a dot the result of the Chiral Quark Model.}
\end{center}
\end{figure*}

We plot the parameters $b$, $M_{QQQ}$ and $E$ as a function of the string tension in Fig.~\ref{sigma}. We vary the value of $\mu_c$ from
0.15 to 0.85 fm$^{-1}$ and leave $a_c$ unchanged so the saturation energy does not change.

We see that for higher string tension values (our value is lower than some determinations) the binding energy will increase.

Our confinement effective potential is
\begin{eqnarray}
        V(r) \sim \sigma \frac{1-e^{-\mu_c r}}{\mu_c}
\end{eqnarray}
We vary the value of $\mu_c$ from 0.15 to 0.85 fm$^{-1}$ and change $a_c$ so that $\sigma$ does not change. This is the same interval
we used when we changed the string tension $\sigma$.
The saturation energy changes as
$\frac{\sigma}{\mu_c}$. Notice that the interaction region is $\sim \sqrt 3 b$ so if $x\equiv \sqrt 3 b \mu_c \ll 1$ the potential in the interacting
region is basically linear. In this calculation we got $x=0.062$ to $x=0.39$ in the charm sector and $x=0.039$ to $x=0.24$ in the bottom sector.
For $\mu_c\to 0$ the potential becomes more linear in the interaction region.

\begin{figure*}[t]
\begin{center}
        \scalebox{0.7}{\includegraphics{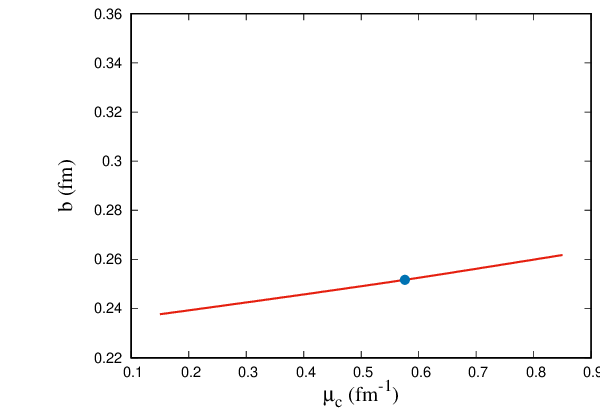}}
        \scalebox{0.7}{\includegraphics{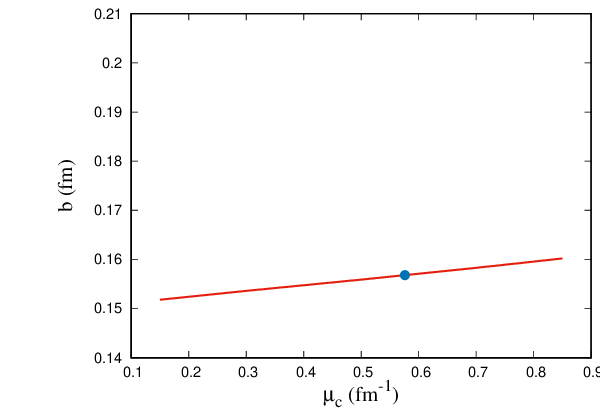}}
        \scalebox{0.7}{\includegraphics{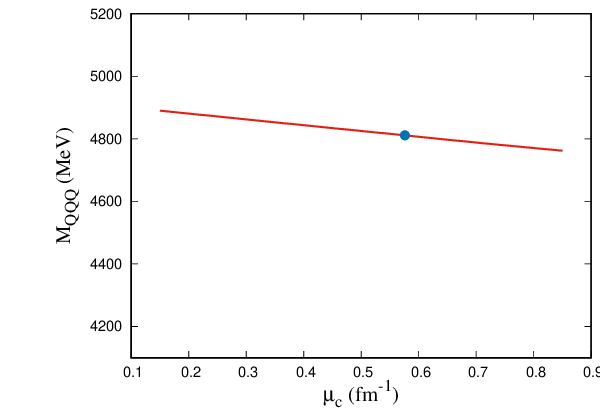}}
        \scalebox{0.7}{\includegraphics{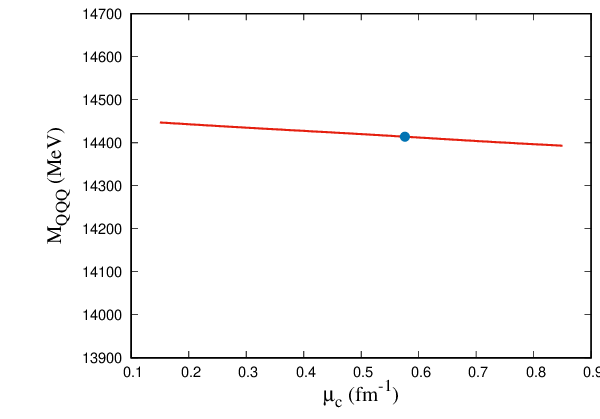}}
        \scalebox{0.7}{\includegraphics{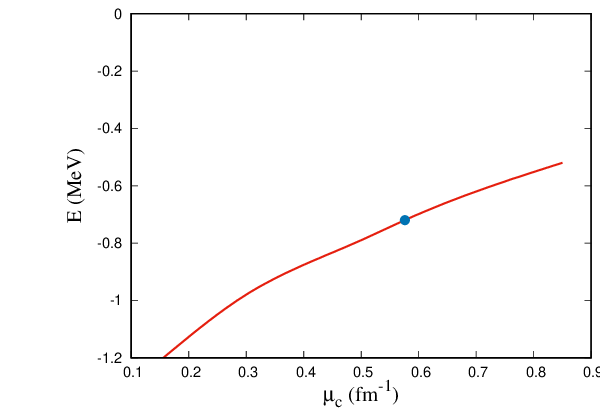}}
        \scalebox{0.7}{\includegraphics{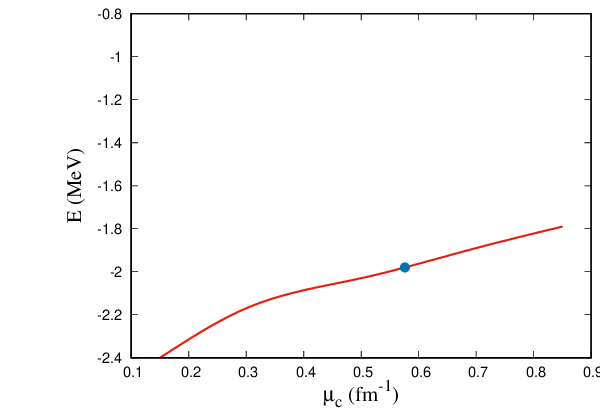}}
        \caption{\label{muc} Dependence of $b$, $M_{QQQ}$ and $E$ on the saturation parameter $\mu_c$ for
$m_q=m_c$ (left) and $m_q=m_b$ (right). We show by a dot the result of the Chiral Quark Model.}
\end{center}
\end{figure*}

The results varying the saturation are shown in Fig.~\ref{muc}. We see that the dependence on the saturation point of the properties of the $\Omega_{QQQ}$, $b$ and $M_{QQQ}$, is smaller than on the
string tension $\sigma$ as one would expect. For the binding energy of the $\Omega_{QQQ}$ dibaryon we see also an smaller dependence.

Notice that when $\mu_c \to 0$ the binding energy increases, so a linear confinement potential should give more binding.

\begin{figure*}[t]
\begin{center}
        \scalebox{0.7}{\includegraphics{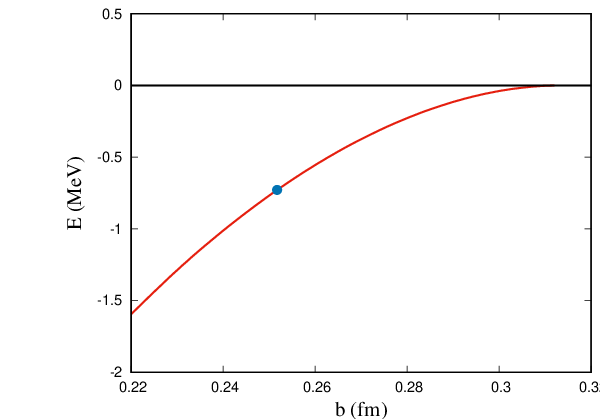}}
        \scalebox{0.7}{\includegraphics{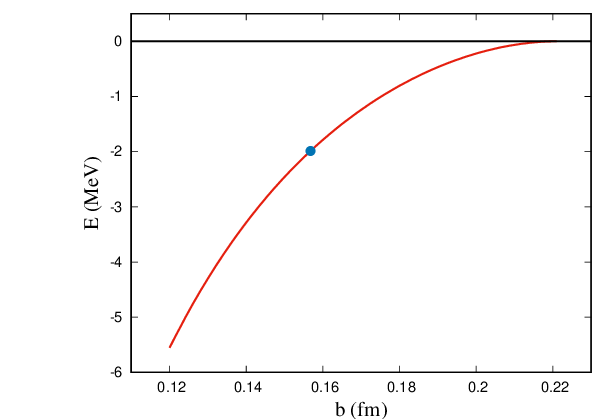}}
	\caption{\label{Evsb} Dependence of $E$ on the size of the baryon given by $b$. Charm sector on the left and bottom sector on the right. We show by a dot the result of the Chiral Quark Model.}
\end{center}
\end{figure*}

Finally we can study the dependence of the binding energy on the size of the baryon. For that we keep all the parameters unchanged and
only vary the parameter $b$ on the RGM calculation. Results are shown in Fig.~\ref{Evsb}.
With bigger sizes we get less binding but it has to be increased much more than the difference between
the sizes of the variational and GEM calculation, which shows that using the exact wave function the system
will still bind. This argument is more robust for the bottom sector but it should also work in the charm.

The result should be seen as an upper bound of the binding energy, since we are using a variational calculation. Also other channels may be involved, but since we are considering the lower energy channel, including
more channels will provide more attraction. We can conclude that the Chiral Quark Model binds the 
$\Omega_{QQQ}\Omega_{QQQ}$ system in both cases, when $Q$ is a bottom or a charm quark. These molecular
states are analogs of two-atom molecules, where the direct interaction is zero for neutral atoms,
as in our model for colorless objects.

\section{Summary}
\label{sec:epilogue}

In this work we have studied the possible existence of fully-heavy dibaryons in the charm and bottom sectors. The main conclusion we found is that, using a wave function which minimizes the mass of the $\Omega_{ccc}$ ($\Omega_{bbb}$) baryons, the six $c$ quarks or the six $b$ quarks can form bound states with $J^P=0^+$ quantum numbers. The binding energy of the charm dibaryon is $E_b=-0.71$ MeV, while in the bottom case the binding energy is slightly higher, $E_b=-1.98$ MeV, which is reasonable due to the highest mass of the bottom quark. The $J^P=0^+$ state corresponds to the coupling of $^1S_0$ and $^5D_0$ partial waves, but with a very small $^5D_0$ component.
No further bound states are found in other partial waves.

\begin{acknowledgments}
This work has been partially funded by EU Horizon 2020 research and innovation program, 
STRONG-2020 project, under grant agreement no. 824093; 
Ministerio Espa\~nol de Ciencia e Innovaci\'on under grant
nos. PID2022-141910NB-I00 and PID2022-140440NB-C22; 
and Junta de Andalucía under contract Nos. PAIDI FQM-370 and PCI+D+i under the title: ”Tecnologías avanzadas para la exploración del universo y sus componentes” (Code AST22-0001).
\end{acknowledgments}

\bibliography{DIB}

\end{document}